\documentclass[runningheads,a4paper]{llncs}

\usepackage{amssymb}
\usepackage{graphicx}

\usepackage{booktabs}
\usepackage{url}
\usepackage{algpseudocode}
\usepackage{amsmath}
\usepackage{amsfonts}
\usepackage[ruled,linesnumbered]{algorithm2e}
\usepackage{multicol}     
\usepackage{multirow}
\usepackage[misc]{ifsym}

\newtheorem{myDef}{Definition}
\urldef{\mailsa}\path|{alfred.hofmann, ursula.barth, ingrid.haas, frank.holzwarth,|
\urldef{\mailsb}\path|anna.kramer, leonie.kunz, christine.reiss, nicole.sator,|
\urldef{\mailsc}\path|erika.siebert-cole, peter.strasser, lncs}@springer.com|    
\newcommand{\keywords}[1]{\par\addvspace\baselineskip
\noindent\keywordname\enspace\ignorespaces#1}

\begin{document}

\mainmatter  

\title{Temporal-Amount Snapshot MultiGraph for Ethereum Transaction Tracking}

\titlerunning{Temporal-Amount Snapshot MultiGraph}

%
%

\author{Yunyi Xie\textsuperscript{1,2}     \and
        Jie Jin\textsuperscript{1,2}       \and
        Jian Zhang\textsuperscript{1,2}    \and
        Shanqing Yu\textsuperscript{1,2}   \and
        Qi Xuan\textsuperscript{1,2,3}\textsuperscript{(\Letter)}
}
\institute{       
\textsuperscript{1} Institute of Cyberspace Security, Zhejiang University of Technology, \\Hangzhou 310023, China \\
\textsuperscript{2} College of Information Engineering, Zhejiang University of Technology, \\Hangzhou 310023, China\\
\textsuperscript{3} PCL Research Center of Networks and Communications, Peng Cheng Laboratory, \\Shenzhen 518000, China\\
\email{xuanqi@zjut.edu.cn}
}
\authorrunning{Y. Xie et al.}

%
%

\toctitle{Lecture Notes in Computer Science}
\tocauthor{Authors' Instructions}
\maketitle

\begin{abstract}
With the wide application of blockchain in the financial field, the rise of various types of cybercrimes has brought great challenges to the security of blockchain. In order to better understand this emerging market and explore more efficient countermeasures for effective supervision, it is imperative to track transactions on blockchain-based systems. Due to the openness of Ethereum, we can easily access the publicly available transaction records, model them as a complex network, and further study the problem of transaction tracking via link prediction, which provides a deeper understanding of Ethereum transactions from a network perspective. Specifically, we introduce an embedding based link prediction framework that is composed of temporal-amount snapshot multigraph (TASMG) and present temporal-amount walk (TAW). By taking the realistic rules and features of transaction networks into consideration, we propose TASMG to model Ethereum transaction records as a temporal-amount network and then present TAW to effectively embed accounts via their transaction records, which integrates temporal and amount information of the proposed network. Experimental results demonstrate the superiority of the proposed framework in learning more informative representations and could be an effective method for transaction tracking.
\end{abstract}

\section{Introduction}
Blockchain is a distributed ledger technology, which can record transactions among peers~\cite{swan2015blockchain}. It could be described as a trusted database that has the characteristics of decentralized, anti-counterfeiting, as well as user tamper-ability and anonymity. With the support of the underlying blockchain technology, blockchain platforms such as Bitcoin and Ethereum also take this opportunity to flourish and become world-renowned new digital currency trading platforms. As the largest public blockchain-based platform enabling smart contracts, Ethereum~\cite{wang2019blockchain} has become a widely used financial application platform and the corresponding cryptocurrency \textit{Ether} is the second-largest cryptocurrency. 

However, along with the rapid development of blockchain technology, various types of cybercrimes have arisen endlessly and thus Ethereum has become a hotbed of various cybercrimes~\cite{yuan2020phishing,wu2020phishers,yuan2020detecting}. Due to the anonymity of the blockchain, criminals attempt to evade supervision and engage in illegal activities by injecting funds into the blockchain system. It's reported that Ethereum has suffered from a variety of scams, such as hacks, phishing, and Ponzi schemes, showing that cybercrimes have become a critical issue in Ethereum~\cite{russon2017ethereum}. In order to create a favorable investment environment and preserve the sustainable development of blockchain-based systems, it's imperative to pay more attention to research in this field for formulating effective supervision. 

This paper focuses on one of the solutions of Ethereum illegal activities, namely transaction tracking. The so-called transaction tracking is to maintain transaction security, identify fraud gangs, trace capital flows, retrieve stolen money, and improve the regulatory system. Furthermore, transaction tracking helps ordinary investors or cryptocurrency companies check whether certain funds or transactions are associated with illegal entities or contaminated by suspicious paths. In a summary, transaction tracking is an effective regulatory measure to prevent crime. The issues of transaction tracking have been widely discussed and many methods have been proposed. However, compared with traditional scenarios, illegal activities on Ethereum behave very differently. Traditional illegal activities generally rely on phishing emails and websites to obtain sensitive information from users~\cite{khonji2013phishing}. As a result, existing methods that aim to detect emails or websites that contain phishing and fraudulent information and thus cannot be directly applied to solve the transaction tracking problem on blockchain platforms like Ethereum.

Thanks to the openness of Ethereum, the available access transaction records that contain rich historical information can be used to study Ethereum matters. Here, we model Ethereum transaction records as a transaction network for further understanding and study transaction tracking on Ethereum from a network perspective, where a node represents an account and each edge represents a particular transaction (containing some unique information such as amount values and timestamps of the transactions). Network embedding~\cite{cui2018survey} could be used to extract meaningful information as to representation features from transaction networks and thus can benefit lots of useful downstream tasks such as node classification~\cite{rossi2012time}, link prediction~\cite{fu2018link,zhang2020hyper}, graph classification~\cite{xuan2019subgraph,zhou2020m}, community detection~\cite{fortunato2010community}, etc. The problem of tracking and predicting transactions on Ethereum transaction networks can be modeled as a link prediction task. 

In this paper, we propose a transaction tracking framework on Ethereum from a network perspective. Firstly, we define a temporal-amount snapshot multigraph~(TASMG) to model the Ethereum transaction records and make the successive snapshots connect to reduce temporal loss. Furthermore, we introduce temporal-amount walk~(TAW) to learn representations of accounts. For each account, the searching strategy depends on the amount transition probability and the temporal transition probability. Various experiments conducted on real-world Ethereum datasets demonstrate that the proposed framework can efficiently learn informative account representation, and solve the problem of transaction tracking.

The main contributions of this paper are summarized as follows. 
\begin{itemize}
\item[$\bullet$] We study the matter of transaction tracking in Ethereum from a network perspective, which provides a deeper understanding of Ethereum transaction records and may contribute to the long-term development of the blockchain.
\item[$\bullet$] We construct temporal-amount snapshot multigraph~(TASMG) to retain Ethereum transactions' temporal and amount properties. Moreover, we propose a temporal-amount walk~(TAW) to effectively embed accounts based on their transactions, which integrates temporal information and amount information of the Ethereum transaction networks. 
\item[$\bullet$] We carry on transaction tracking experiments on realistic Ethereum datasets, which demonstrate that our proposed framework significantly outperforms the baselines in transaction tracking. More precisely, our method combined with Ethereum datasets' properties can be superior to general methods. 
\end{itemize}

The remainder of this work is organized as follows. In Section~\ref{sec:RelatedWork}, we give a review of related work on blockchain transaction data analysis and summarize the related work on network embedding and link prediction. In Section~\ref{sec:Methodology}, we demonstrate the framework for Ethereum transaction tracking. In Section~\ref{sec:Experiment}, we conduct extensive experiments on real-world Ethereum datasets and compare our method with several network embedding techniques and similarity-based link prediction methods. Finally, we conclude our work in Section~\ref{sec:Conclusion}.

\section{Related Work} \label{sec:RelatedWork}
\subsection{Blockchain Transaction Analysis and Mining}
In the traditional financial scenario, transaction records are sensitive information, which is usually private for security. Thanks to the openness of blockchain with publicly accessible transaction records, researchers can independently access Ethereum transaction records, which brings unprecedented opportunities for blockchain analysis. The study of blockchain transaction records has recently attracted considerable attention from different applications like graph analysis, price prediction, and anti market manipulation. 

Wu et al.~\cite{wu2020detecting} proposed the concept of \textit{Attributed Temporal Heterogeneous Motifs} and further addressed the issue of mixing detection using a detection model. Recently, they presented a network embedding model named \textit{trans2vec} for transaction networks, which incorporates the transaction amount values and timestamps. It is worth noting that the model makes contributes to phishing detection~\cite{wu2020phishers,yuan2020detecting} and can be applied to other similar scenarios on transaction networks. Current researches on the blockchain mainly focus on cybercrimes, we argue that such preference may promote short-term security but may not conducive to the long-term development of the blockchain. In this paper, we consider the prediction of transactions and try to explore a deeper understanding of Ethereum transactions from a network perspective.

\subsection{Network Embedding and Link Prediction}
Network embedding has received much attention over the past decades~\cite{cui2018survey}. Such methods intuitively focus on transforming each node into a low-dimensional vector, in which the structural information and topology properties of nodes are preserved as much as possible. Link prediction~\cite{lu2011link} aims at estimating the likelihood of the existence of links between nodes based on the current network information. Existing link prediction methods can be classified into several categories, e.g., similarity-based algorithms, maximum likelihood algorithms, and probabilistic models and most of the existing network embedding techniques can be applied to link prediction. 

One of the earliest efforts in network embedding is to combine random walk based methods with Skip-Gram~\cite{mikolov2013efficient} model to learn node representation, where DeepWalk~\cite{perozzi2014deepwalk} and Node2vec~\cite{grover2016node2vec} are two classic examples using random walks on networks to obtain node representation. Tang et al.~\cite{tang2015line} proposed another successful network embedding model LINE, which designs the objective function, optimizes its first-order and second-order proximity, and performs the optimizations by stochastic gradient descent with edge sampling. Factorization based network embedding represents the connections between nodes in the form of a matrix and obtains the node embedding by factorizing this matrix. Graph Factorization~(GF)~\cite{ahmed2013distributed} and HOPE~\cite{ou2016asymmetric} are two of the most notable factorization based methods. GCN based autoencoder~\cite{kipf2016variational}, e.g., GAE and VGAE, has been widely used, which aims to minimize the reconstruction error of the output and input by its encoder and decoder. 

Considering the high expressiveness and learning ability of random walk based methods and the diversity of Ethereum transaction, we model general transaction network as a temporal-amount snapshot multigraph~(TASMG), which combines transaction temporal and amount information. Furthermore, we implement a random walk using specific search strategies, namely temporal-amount walk~(TAW), on the proposed TASMG. In this algorithm, each node has its unique search strategy, reflecting the transaction amount and temporal properties simultaneously. 

\section{Methodology} \label{sec:Methodology}
\subsection{Basic definition}\label{sec:definition}
\begin{figure}[!t]
  \centering
  \includegraphics[width=0.85\linewidth]{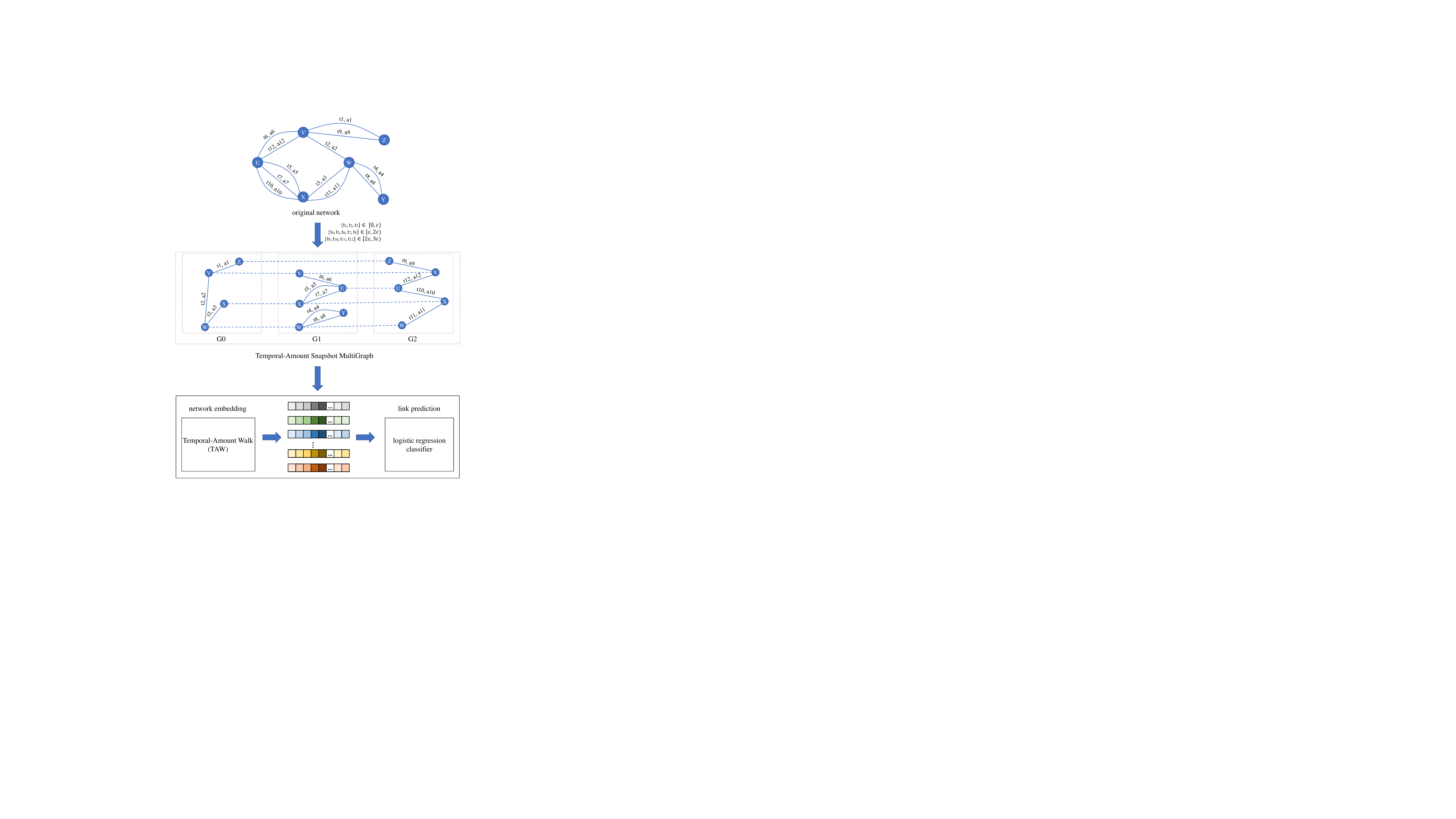}
  \caption{The detailed framework for Ethereum transaction tracking.}
  \label{fig:process}
\end{figure}

In general, the Ethereum transaction records can be modeled as a network $G=(V, E)$, where $V$ denotes the set of accounts~(nodes) and $E$ represents the transaction records~(edges) with transaction temporal and amount information. According to the given time interval $\epsilon$, the transaction network $G$ can be divided into several snapshots $\{G_{1},G_{2},G_{3},\cdots\}$. In order to capture the changing tendency of accounts' transaction behavior, it is crucial to consider not only the snapshot of the current time but also the snapshot of nearby time. Hence, we define TASMG to formulate our solution and further propose TAW to capture the temporal and amount properties of each account in TASMG. Fig.~\ref{fig:process} demonstrates the main steps of the proposed framework, including network construction, network embedding, and transaction tracking.

\begin{myDef}[temporal-amount snapshot multigraph~(TASMG)] Graph $G=(V,E)$ can be divided into several snapshots $\{G_{0}, G_{1}, G_{2}, \cdots\}$ according to time span $\epsilon$, where $G_{t}=(V_{t},E_{t})$. Let $V_{t}$ and $E_{t}$ be sets of nodes and edges of snapshot $G_{t}$ respectively, and they are active between the timespan $[t\epsilon,(t+1)\epsilon)$, where time order $t \in \{0, 1, 2, \cdots\}$. Each edge is unique and is represented as $e = (u, v, w, t)$, for $\forall e \in E$, $Src(e) = u$, $Dst(e) = v$, $W(e) = w$, $T(e) = t$, where $u$ is the source node, $v$ is the target node, $w$ is the weight~(transaction amount value) and $t$ is the time accessibility. All snapshots are sorted by time order $t$~(ascending), and self-connections could be established when and only when a node existed in successive snapshots. 
\end{myDef}

Self-connections in TASMG can make the random walk traverse across successive snapshots in non-decreasing order of edge's temporal information, which can capture the correlation between different snapshots and thus may result in more informative embeddings. For simplicity, we provide time accessibility for the edge between each pair-wise nodes and every two time slices linked nodes. Let $\eta_{+} : \mathbb{R} \to \mathbb{Z}^{+}$ be a function that maps each node to an index based on the time order $t$, i.e., for a given node $u$ in snapshot $G_{i}$, we have $\eta_{+}(u) = i$. Therefore, we can design a mapping function for each edge $e$ in TASMG: $T(e) = \eta_{+}(v) - \eta_{+}(u) \in \{-1, 0, 1\}$, where $v$ is the target node and $u$ is the source node, to define the time accessibility of $v$ from $u$, i.e., $v$ is time accessible from $u$ if and only if the corresponding $T(e)\geq{0}$. Then, we define \textit{successive edges} as follows.

\begin{myDef}[Accessible Edge] Given a TASMG ${G=(V,E)}$, the set of accessible edges for a node $v$ is defined as:
  \begin{equation*}\label{eq:successiveedges}
  L_{t}(v) = \{e\,|\,Src(e)=v, T(e) \geq 0\}
  \end{equation*}
\end{myDef}

In the proposed network TASMG, a valid temporal walk is composed of accessible edges, i.e., a sequence of nodes connected by edges with non-decreasing $T(e)$. Next, we will introduce different sampling strategies by formulating the selection probability for each accessible edge $e \in L_{t}(v)$. 

\subsection{Temporal biased walk}\label{sec:temporalwalk}
Based on the definitions in Sec.~\ref{sec:definition}, we design a sampling strategy to choose accessible edges both in temporal and amount domains. 

\subsubsection*{Temporal Transition Probability:} When we divide the whole transaction network into different snapshots based on time span, each snapshot represents a part of the network structure and the gradual change of time slice reflects the evolution process of network. Ignoring the correlation information that exists between two snapshots at consecutive time steps may cause the loss of temporal information. Hence, we propose temporal transition probability to capture node's behavior changes in different snapshots. In this case, the probability of selecting each edge $e \in L_{t}(c)$ can be given as:
\begin{equation}
  P_{T}(e) = \frac{\psi_{T}(e)}{\sum_{e^{'} \in L_{t}(c)}\psi_{T}(e^{'})}
\end{equation}
where $\psi_{T}(e)$ is expressed as
\begin{equation}
  \psi_{T}(e) =  \left\{
\begin{array}{rcl} 
  \alpha,     &     &T(e) > 0\\
  1-\alpha,   &     &T(e) = 0
\end{array} \right. 
\end{equation}
Here, the temporal bias $\alpha$ $(0.1 \le \alpha \le 0.9)$ decides whether the temporal walk resides on the current snapshot or transfers to the next. 

\subsubsection*{Amount Transition Probability:} Apart from transaction time, the transaction amount value of edges also plays an essential role in transaction networks. In the following, we present unbiased, biased, and linear strategies from the amount domain, respectively.
\begin{itemize}
  \item[$\bullet$] \textbf{Amount Unbiased Sampling~(AUS).} This is the default setting in the amount domain, which assumes that each successive edge $e \in L_{t}(c)$ of node $c$ has the same probability to be selected:
  \begin{equation}
    P_{A}(e) = \frac{1}{L_{t}(c)}
  \end{equation}
  \item[$\bullet$] \textbf{Amount Biased Sampling~(ABS).} The amount value of each transaction indicates the importance of the interaction between the two accounts involved. In most cases, a higher value of transaction amount implies that there is a greater connection between the two accounts. Thus each edge $e \in L_{t}(c)$ can be assigned the selection probability:
  \begin{equation}
    P_{A}(e) = \frac{W(e)}{\sum_{e^{'} \in L_{t}(c)}W(e^{'})}
  \end{equation}
  where $W(e)$ is the transaction amount value between node $c$ and its temporal neighbor $x$. 
  \item[$\bullet$] \textbf{Amount Linear Sampling~(ALS).} In order to avoid the extreme situation that the edge with a small amount value will never be sampled or the edge with a large amount value will always be sampled, we consider a linear mapping function to weaken the effects of the edge's transaction amount value. Thus we have: 
  \begin{equation}
    P_{A}(e) = \frac{\omega_{+}(W(e))}{\sum_{e^{'} \in L_{t}(c)}\omega_{+}(W(e^{'}))}
  \end{equation}
  with $\omega_{+}(v)$ mapping the amount value of edge to an ascending ranking. In other words, $\omega_{+}(v)$ maps each edge's transaction amount value to an index with $\omega_{+}(e) = 1$ for the smallest transaction amount value. 
\end{itemize}

\subsubsection*{Joint Transition Probability:} Furthermore, we normalize the aforementioned temporal transition probability and amount transition probability, and then combine them as one. We set the unnormalized transition probability to $P(e)$ and then normalize it to the final transition probability for each edge $e \in L_{t}(c)$, where 
\begin{equation}
  P(e) = P_{T}(e)\,P_{A}(e)
\end{equation}

\subsection{Learning network embeddings}
Our goal is to obtain a mapping function $f:V \to \mathbb {R}^{d}$, which maps a given node to a $d$-dimensional representation. For a node $v \in V$, let $N(v)$ denotes the set of temporal neighbors that are generated according to the sampling strategy, and $f_{t}(v)$ is the representation of node $v$ in snapshot $G_{t}$. Our objective function aims to maximize the log-probability of observing $N(v)$ and historical embedding $f_{t}(v)$ for the node $v$ conditioned on its representation:
\begin{equation}\label{eq:1}
\max_{f}\sum_{v \in V}\log(Pr(N(v),f_{t}(v)\,|\,f(v)))
\end{equation}

According to the conditional independence assumption in the Skip-Gram model, we factorize the formula: 
\begin{equation}
\label{eq:2}
\begin{aligned}
&\log(Pr(N(v),f_{t}(v)\,|\,f(v))) \\
=&\log(\prod_{u_{i} \in N(v)}Pr(u_{i}\,|\,f(v))) + \log(Pr(f_{t}(v)\,|\,f(v)))
\end{aligned}
\end{equation}

Based on the network analysis, we can see that the likelihood of observing a source node is independent of observing any other and the definition of neighborhood nodes is symmetric. Therefore, we factorize the likelihood of observing temporal neighbors and model the likelihood of every source-neighborhood node pair as a softmax unit. The conditional probability of observing a node $u_{i}$ given the learned representation $f(v)$ can be transformed as follows:
\begin{equation}
\label{eq:3}
Pr(u_{i}\,|\,f(v))=\frac{\exp(f(u_{i})\,f(v))}{\sum_{n \in V}\exp(f(n)\,f(v))}
\end{equation}
where $u_{i} \in N(v)$ is the $i$th near-neighbor of node $v$. With the above hypothesis, the objective function in Eq.~(\ref{eq:1}) simplifies to:
\begin{equation}
\label{eq:4}
\begin{aligned}
\max_{f}\sum_{v \in V}\log(\prod_{u_{i} \in N(v)} &\frac{\exp(f(u_{i})\,f(v))}{\sum_{n \in V}\exp(f(n)\,f(v))}) \\
&+\log(Pr(f_{t}(v)\,|\,f(v)))
\end{aligned}
\end{equation}

\section{Experimental Results} \label{sec:Experiment}
\subsection{Dataset}
\begin{figure}[!t]
  \centering
  \includegraphics[width=0.65\linewidth]{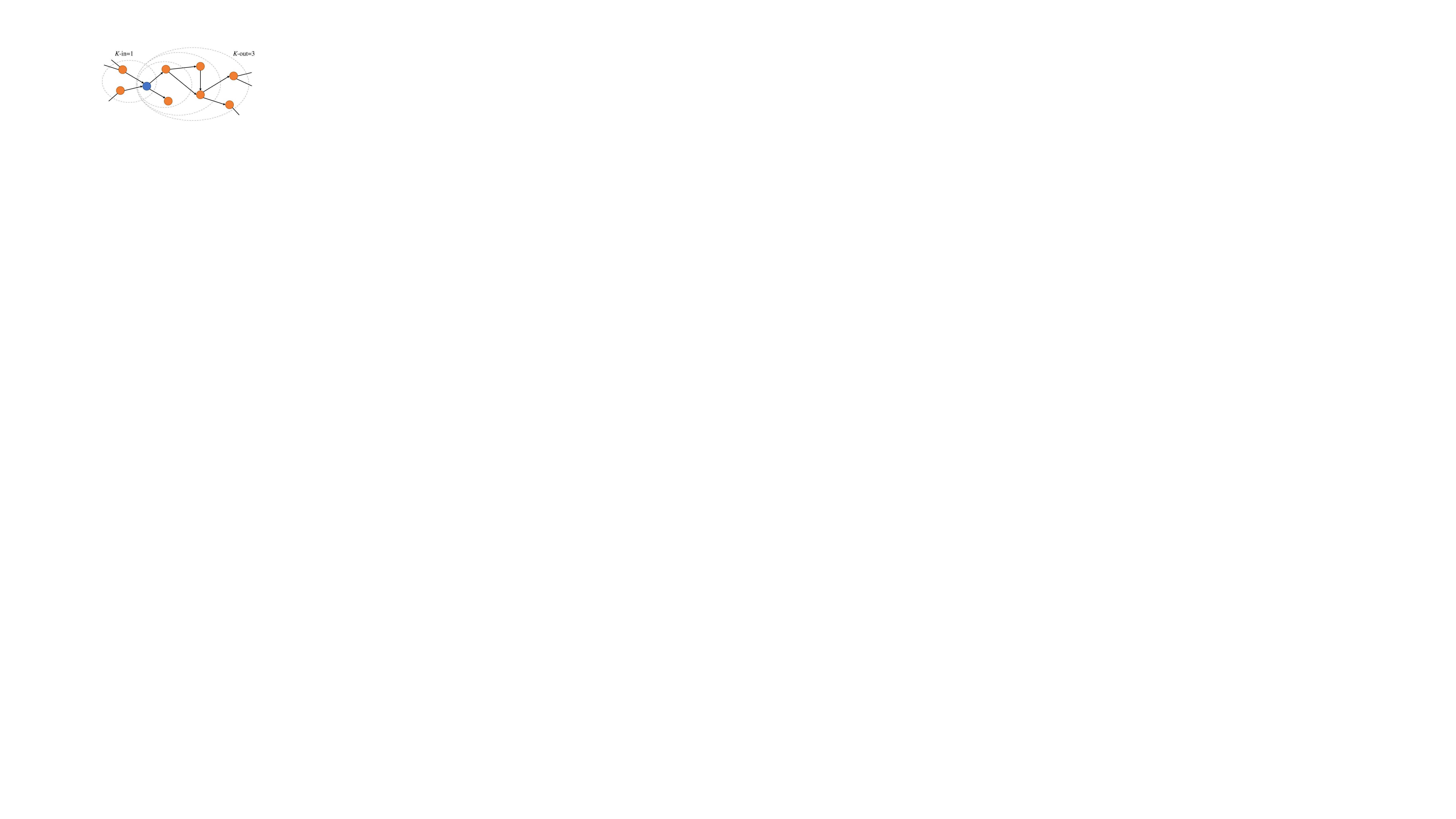}
  \caption{Schematic illustration of $K$-order subgraph. }
  \label{fig:kordersubgraph}
\end{figure}

As the largest public blockchain-based platform, Ethereum's~\cite{wood2014ethereum} transaction records are completely public and researchers can easily obtain the objective account's historical transaction data through the API of Etherscan~(\url{etherscan.io}). Due to the size of the total transaction records is extremely large~\cite{chen2018understanding}, we ascertain a number of target accounts and then obtain their transactions from Ethereum transaction records to make subgraphs for subsequent experiments. As shown in Fig.~\ref{fig:kordersubgraph}, we randomly sample a centered account to obtain its local structure information and then extract $K$-order subgraph~\cite{lin2020modeling}. $K$-in and $K$-out are two parameters to control the depth of sampling inward and outward from the center, respectively. 

In the following experiments, we collect three subgraphs for transaction tracking with different sizes from the whole Ethereum transaction records, i.e., randomly select different center accounts and collect three subgraphs with $K$-in = 1, $K$-out = 3. A summary of these networks is listed in Table~\ref{tab:dataset}.
\begin{table*}[!t]
  \setlength{\tabcolsep}{6.5mm}
  \centering
  \renewcommand{\arraystretch}{1.2}
  \setlength{\abovecaptionskip}{0pt}
  \setlength{\belowcaptionskip}{5pt}
  \caption{Basic topological features of the three subgraphs. $|V|$ and $|E|$ are the numbers of nodes and edges, respectively, $\langle K \rangle$ is the average degree, and $\langle C \rangle$ is the average clustering coefficient.}
  \begin{tabular}{ccccc}
  \toprule
              & $|V|$      & $|E|$    & $\langle K\rangle$  & $\langle C\rangle$ \\ \midrule
  EthereumG1  & 2100       & 6995     & 6.662               & 0.211   \\
  EthereumG2  & 5762       & 9098     & 3.158               & 0.112   \\
  EthereumG3  & 10269      & 28431    & 5.537               & 0.147   \\ \bottomrule
  \end{tabular}
  \label{tab:dataset}
\end{table*}

\subsection{Baselines and Experimental Setup}
It is assumed that node pairs with higher similarity values are more likely to have interactions, and thus we calculate the similarity value for each node pair, e.g., Common Neighbors~(CN), Adamic–Adar Index~(AA), Resource Allocation Index~(RA). Furthermore, we compare the performance of TAW with several baseline methods including GF, HOPE, GAE, VGAE, LINE, DeepWalk, and Node2vec. GF and HOPE are two of the most notable factorization based embedding methods, while GAE and VGAE are the GCN based autoencoder models. LINE preserves the first-order and second-order proximity between nodes and in this work, the final representation for each node is created by second-order representations. DeepWalk and Node2vec belong to the same category of Skip-Gram based methods. The parameters for both the node sampling and optimization steps of the two random walk based embedding methods, including ours, are set exactly the same: number of walks per node $w = 10$, the length of walk $l = 80$, and the size of context window $k = 5$. Since Node2vec requires the in-out and return hyperparameters, the ranges of its hyper parameters in this paper are set to $p,q\in\{0.5, 1, 2\}$. For our proposed method TAW, we take one month as the time span to construct TASMG for each dataset and temporal bias $\alpha$ varies in $[0.1,0.9]$. 

For all methods, the number of dimensions of output vector representations $d$ is set to 128. After learning embedding for each node, we use hadamard $\footnote{$[f(u) \cdot f(v)]_{i}=f_{i}(u) * f_{i}(v)$}$ operation on the learned embedding vectors of pair-wise nodes to compute the feature vector for the corresponding edge. And we train a one-vs-rest logistic regression classifier to classify the links in the test set. Experiments are repeated for 5 random seed initializations and we report the average performance. 

\subsection{Metrics}
\begin{itemize}
  \item[$\bullet$] \textbf{AUC:} It can be interpreted as the probability that a randomly chosen missing link is given a higher score than a randomly chosen nonexistent link. If among $n$ independent comparisons, there are $n^{'}$ times that the missing link gets a higher score and $n^{''}$ times they get the same score, the AUC value is
  \begin{equation*}
    \rm AUC = \frac{n^{'}+0.5n^{''}}{n}
  \end{equation*}
  If all the scores are generated from an independent and identical distribution, the AUC value should be about 0.5. The degree to which the value exceeds 0.5 indicates how much better the algorithm performs than pure chance.
  \item[$\bullet$] \textbf{Precision:} Given the ranking of the non-observed links, the precision is defined as the ratio of relevant items selected to the number of items selected, which indicates how many predictions are accurate, from the perspective of prediction results. That is to say, if we take the top-$L$ links as the predicted ones, among which $L_{r}$ links are right, then the Precision equals $L_{r}/L$. Clearly, higher precision means higher prediction accuracy.
\end{itemize}

\subsection{Results and Discussions}
We treat the transaction tracking problem as a link prediction task in Ethereum, which aims to predict the occurrence of links in a given network based on observed information. Before the experiments, we hide a certain fraction of accounts' connections in the transaction network, and our goal is to trace these missing connections via the use of all remaining information to predict accounts' transaction records. We first randomly hide 20\% of links in the original network as positive samples of the training set and use the remaining to train all methods. Then we stochastically sample an equal number of node pairs with no link as negative samples of the training set. The test set consists of two parts, one of which contains all the hidden edges as the positive samples, and the other contains the unconnected pairwise nodes sampled randomly as the negative samples. 

\begin{table}[!t]
  \setlength{\tabcolsep}{3mm}
  \centering
  \renewcommand{\arraystretch}{1.2}
  \setlength{\abovecaptionskip}{0pt}
  \setlength{\belowcaptionskip}{5pt}
  \caption{Performances of different methods for transaction tracking.}
  \begin{tabular}{ccccccc}
  \toprule
  \multirow{2}{*}{Metrics} & \multicolumn{2}{c}{EthereumG1}  & \multicolumn{2}{c}{EthereumG2}    & \multicolumn{2}{c}{EthereumG3}    \\ \cmidrule(l){2-7}
          & AP       & AUC       & AP       & AUC       & AP       & AUC             \\ \midrule
  CN       & 0.6907   & 0.6848    & 0.5134   & 0.4921    & 0.6826   & 0.6881          \\
  AA       & 0.7367   & 0.7002    & 0.5909   & 0.5099    & 0.6979   & 0.6912          \\
  RA       & 0.7378   & 0.7007    & 0.5909   & 0.5099    & 0.6986   & 0.6914          \\
  Jaccard  & 0.5088   & 0.6097    & 0.4523   & 0.4588    & 0.6234   & 0.6800          \\
  GF       & 0.7827   & 0.6821    & 0.7377   & 0.7050    & 0.7946   & 0.6662          \\
  HOPE     & 0.7698   & 0.6578    & 0.8300   & 0.7580    & 0.8714   & 0.8089          \\
  LINE     & 0.7761   & 0.7521    & 0.8627   & 0.8237    & 0.6371   & 0.6370          \\
  DeepWalk & 0.6159   & 0.6637    & 0.6138   & 0.6307    & 0.7755   & 0.8024          \\
  Node2vec & 0.6877   & 0.7149    & 0.6939   & 0.6990    & 0.8239   & 0.8501          \\
  GAE      & 0.7911   & 0.6752    & 0.5828   & 0.3729    & 0.8703   & 0.7885          \\
  VGAE     & 0.8179   & 0.7184    & 0.6683   & 0.4719    & 0.8934   & 0.8278          \\
  TAW      & \textbf{0.8774} & \textbf{0.8819} & \textbf{0.9180} & \textbf{0.9134} & \textbf{0.9115}    & \textbf{0.9108} \\ \bottomrule
  \end{tabular}
  \label{tab:lp1}%
\end{table}

\textbf{Transaction Tracking Performance.} The experimental results of transaction tracking are given in Table~\ref{tab:lp1}. We find that similarity indices, i.e., CN, Jaccard, RA and AA, perform poorly among all methods. These similarity indices could only extract incomplete information, which is hard for mining deep structure information of transaction networks. We further observe that the sparser of network, the harder it for GCN-based methods to predict the appearance of links. GF, HOPE, and LINE achieve poor performance on most networks, indicating that preserving higher-order proximity is not conducive to predicting unobserved transactions. Here, the random walk based methods outperform other methods, indicating that random walks are especially useful when approximate node centrality and similarity in Ethereum transaction networks. However, it is still not as good as the results of TAW, for the reason that our proposed transaction tracking framework is more meticulous in the way of generating TASMG and extracting more precise features. Overall, the evaluation indicates that TAW achieves clear performance gains over the baselines for transaction tracking problem, which is reasonable since our method with flexible walking strategies is able to learn the similarity between nodes more effectively. Moreover, the outstanding performance of our proposed transaction tracking framework also suggests that TASMG, which better combines temporal and amount properties, can indeed uncover richer valuable information from the different dimensions of Ethereum transaction records. It is worth noting that our TASMG is quite general, i.e., many other random walk based approaches can also be generalized by using our proposed TASMG and can be applied in many other applications beyond Ethereum. 

\begin{figure}[!t]
  \centering
  \includegraphics[width=\linewidth]{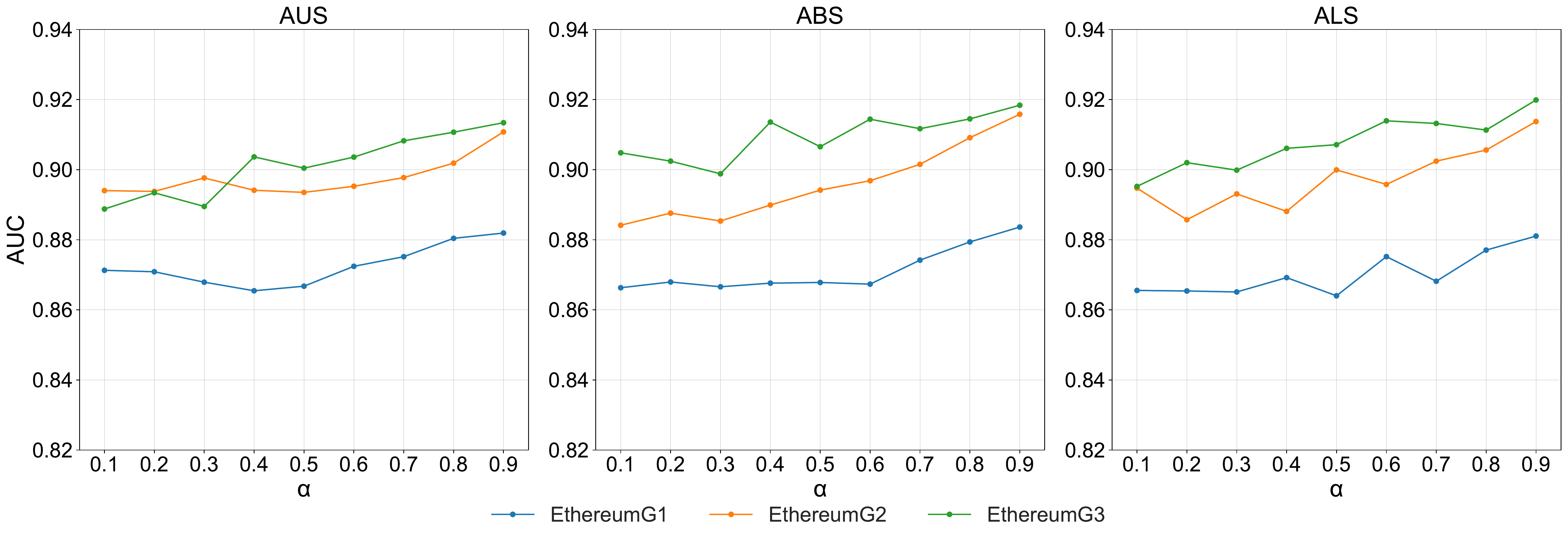}
  \caption{Performance in terms of AUC under temporal bias $\alpha$ which varies in $[0.1, 0.9]$.}
  \label{fig:parameter}
\end{figure}

\textbf{Parameter Sensitivity.} Table~\ref{tab:lp2} shows the difference between different amount sampling strategies. Interestingly, using a biased strategy seems to improve slightly on the tested datasets, which implies that the probability of transaction occurrence is positively correlated with the amount value of the transaction between the two accounts. We thus can infer that the larger transaction amount value of the two accounts, the closer relationship between the two. Furthermore, we  focus on the effects of temporal bias $\alpha$ on the proposed framework, since this parameter is newly introduced in our method. In order to evaluate how temporal bias $\alpha$ could impact the transaction tracking performance, we gradually vary the temporal bias $\alpha$ in $[0.1, 0.9]$. The results are shown in Fig.~\ref{fig:parameter}, where we can observe that, overall, as the increasing of temporal bias $\alpha$, our framework has achieved better performance for transaction tracking. 

\begin{table}[!t]
  \setlength{\tabcolsep}{3mm}
  \centering
  \renewcommand{\arraystretch}{1.2}
  \setlength{\abovecaptionskip}{0pt}
  \setlength{\belowcaptionskip}{5pt}
  \caption{Performances of different amount sampling strategies.}
  \begin{tabular}{ccccccc}
  \toprule
  \multirow{2}{*}{Metrics} & \multicolumn{2}{c}{EthereumG1}  & \multicolumn{2}{c}{EthereumG2}    & \multicolumn{2}{c}{EthereumG3}    \\ \cmidrule(l){2-7}
          & AP   & AUC      & AP       & AUC      & AP       & AUC             \\ \midrule
  AUS & 0.8774   & 0.8819   & 0.9180   & 0.9134   & 0.9115   & 0.9108   \\
  ABS & \textbf{0.8852}   & \textbf{0.8836}   & \textbf{0.9287}   & 0.9184   & \textbf{0.9189}   & \textbf{0.9158}   \\
  ALS & 0.8798   & 0.8811   & 0.9257   & \textbf{0.9199}   & 0.9150   & 0.9137   \\ \bottomrule
  \end{tabular}
  \label{tab:lp2}%
\end{table}

\section{Conclusion} \label{sec:Conclusion}
In this paper, we model a general transaction network as a temporal-amount snapshot multigraph~(TASMG), which provides a novel network perspective for a deeper understanding of Ethereum transactions. Furthermore, we implement a random walk using specific search strategies to characterize accounts' behaviors on the proposed TASMG, namely temporal-amount walk~(TAW). Experimental results demonstrate the effectiveness of our proposed transaction tracking framework, and indicate that TASMG has the potential to extract more information from Ethereum transaction records. Though Ethereum transaction records are publicly available, it's still relatively unexplored till now. For future work, we plan to apply deep learning methods to expand our methods and further extend the current framework to analyze more illegal activities on Ethereum and create a safe trading environment for Ethereum.

\subsubsection*{Acknowledgments.} This work was partially supported by the National Key R\&D Program of China under Grant No. 2020YFB1006104, by the National Natural Science Foundation of China under Grant No. 61973273, and by the Zhejiang Provincial Natural Science Foundation of China under Grant No. LR19F030001.

\end{document}